# Magnetic memory and spontaneous vortices in a van der Waals superconductor

Eylon Persky[1,2,*], Anders V. Bjørlig[1,2], Irena Feldman[3], Avior Almoalem[3], Ehud Altman[4,5], Erez Berg[6], Itamar Kimchi[7], Jonathan Ruhman[1], Amit Kanigel[3] and Beena Kalisky[1,2,*]

1. Department of Physics, Bar Ilan University, Ramat Gan 5920002, Israel
2. Bar Ilan Institute of Nanotechnology and Advanced Materials, Bar Ilan University, Ramat Gan 5920002, Israel
3. Department of Physics, Technion–Israel Institute of Technology, Haifa 32000, Israel
4. Department of Physics, University of California, Berkeley, California 94720, USA
5. Materials Sciences Division, Lawrence Berkeley National Laboratory, Berkeley, California 94720, USA
6. Department of Condensed Matter Physics, Weizmann Institute of Science, Rehovot 76100, Israel
7. School of Physics, Georgia Institute of Technology, Atlanta, Georgia 30332, USA

## Abstract

Doped Mott insulators exhibit some of the most intriguing quantum phases of matter, including quantum spin-liquids, unconventional superconductors, and non-Fermi liquid metals[1–3]. Such phases often arise when itinerant electrons are close to a Mott insulating state, and thus experience strong spatial correlations[4,5]. Proximity between different layers of van der Waals heterostructures naturally realizes a platform for experimentally studying the relationship between localized, correlated electrons and itinerant electrons. Here, we explore this relationship by studying the magnetic landscape of 4Hb-TaS$_2$, which realizes an alternate stack of a candidate spin liquid and a superconductor[6]. We report on a spontaneous vortex phase whose vortex density can be trained in the normal state. We show that time reversal symmetry is broken above T$_c$, indicating the presence of a magnetic phase independent of the superconductor. Strikingly, this phase does not generate detectable magnetic signals. We use scanning superconducting quantum interference device (SQUID) microscopy to show that it is incompatible with ferromagnetic ordering. The discovery of this new form of hidden magnetism illustrates how combining superconductivity with a strongly correlated system can lead to new, unexpected physics.

## Main text

In Van der Waals heterostructures, proximity between layers of different materials can be exploited to generate new states of matter[7–9], or to use one layer in order to probe the properties of the other[10,11]. Indeed, correlated insulating, superconducting, nematic and time-reversal symmetry broken states emerge when uncorrelated electronic systems are stacked together[12–16]. Heterostructures involving strongly correlated systems as their constituents therefore hold promise to realize new phases or find new coupling mechanisms between the layers. A particularly interesting set of ground states to pair is a superconductor and a Mott insulator. Unconventional superconductivity often emerges when a Mott insulator is destroyed by doping[2,3,12], but how these two phases interact when stacked as individual building blocks remains to be explored.

This combination is naturally realized in 4Hb-TaS$_2$, in which two 2D structures of TaS$_2$, octahedral (1T) and trigonal prismatic (2H), are alternatingly stacked[17]. In bulk form, 2H-TaS$_2$ is a superconductor with T$_c$ = 0.7 K [18], while 1T-TaS$_2$ is a correlated insulator[19] with electrons localized on a triangular lattice, predicted to host a quantum spin liquid ground state[20,21]. This suggestion is supported by muon spin relaxation experiments[22–24] and nuclear magnetic resonance measurements[23,25] which show an absence of long-range magnetic order. Furthermore, although resistivity clearly shows an insulating behavior, residual specific heat[22,24] and thermal conductivity[26], and scanning tunneling microscopy measurements[27,28] indicate gapless neutral excitations. Combined, these results point towards a gapless spin liquid ground state in bulk 1T-TaS$_2$.

The 4Hb compound therefore realizes a unique system, where layers of a superconductor and a candidate SL are alternately stacked (Figure 1a,b). Recent scanning tunneling microscopy studies on 1T/1H heterostructures of TaS$_2$ and TaSe$_2$ revealed that the localized and itinerant electrons in the 1T and 1H layers interact through a Kondo coupling[27,28]. Such an interaction can melt the 1T insulator into a metal through a mean-field hybridization with the 1H layers, akin to doping the spin liquid, which allows for a variety of new ground states to emerge. Indeed, the superconducting phase of 4Hb-TaS$_2$ is anomalous, compared to the bulk 2H polymorph: its critical temperature is elevated to ~2.7 K and it breaks time-reversal symmetry (TRSB) below T$_c$ [6]. The fate of the 1T layers is particularly interesting. They may order magnetically[29,30], or become metallic due to charge transfer to the 1H layers, in which case Mott physics is no longer relevant. On the other hand, they may remain in a gapless spin-liquid state or assume different spin-liquid order at intermediate coupling. Recent numerical studies suggest that a chiral spin liquid is likely at intermediate coupling[31,32]. Moreover, when such a CSL ground state is lightly doped, it may give way to chiral superconductivity and chiral metallicity[33].

Here, we investigate this system using scanning SQUID microscopy. We study the magnetic landscape of the sample in both the superconducting and normal phases, and in the presence of small external magnetic fields. In the superconducting phase, the TRSB is manifested in a spontaneous vortex phase. The magnetic and thermal history of the spontaneous vortex state reveal that TRSB persists up to 3.6 K, above the superconducting transition. Surprisingly, we show that this TRSB order is "hidden" above T$_c$, as it does not produce detectable magnetic signatures in the normal state. These results are inconsistent with conventional ferromagnetic ordering or with conventional coupling between magnetism and superconductivity. We propose a CSL in the 1T layers as a candidate TRSB phase. Our results demonstrate how proximity between strongly correlated constituents of a Van der Waals heterostructure can give rise to novel electronic phases.

### Spontaneous vortex phase and magnetic memory

We studied two 4Hb-TaS$_2$ single crystals (Methods). To study the superconductivity, we used scanning SQUID magnetometry and susceptometry (Methods). The local susceptometry measurements revealed a homogeneous diamagnetic response over our field of view both at our base temperature, 1.7 K, and closer to the critical temperature, T$_c$ = 2.7 K (Extended Data Figure 1a). Here, we define T$_c$ as the temperature at which the local diamagnetic response disappears. All samples showed a sharp transition around 2.7 K, and at 2.8 K the diamagnetic response disappeared, leaving no traces within our sensitivity. Furthermore, vortices appearing in the DC magnetometry images below $T_c$ disappeared at 2.8 K. Combined, these data show that the entire field of view was normal at 2.8 K (Extended Data Figure 1b). Scanning tunneling microscopy measurements[34], Specific heat measurements[6], as well as global magnetometry and resistivity measurements (Extended Data Figure 1d,e) also show a sharp superconducting transition at 2.7 K.

In type II superconductors, the vortex density is determined by the magnetic field present when the superconductor is cooled through its critical temperature. By imaging the distribution of individual vortices in the superconducting phase, we infer the two-dimensional vortex density, $n_v$, and the corresponding average magnetic field, $B = n_v \Phi_0$. Figure 1 summarizes the main evidence for hidden magnetization in 4Hb-TaS$_2$. First, when zero field cooling (ZFC) the sample from 4 K to 1.7 K, we observed a single vortex in a 95 μm by 75 μm region (Figure 1c), indicating that the external field was < 5 mG (Methods). Field cooling (FC) the sample in an external field of 1 G generated a denser distribution of vortices, with 287 of them in the same region. Surprisingly, a subsequent ZFC resulted in 32 vortices (Figure 1d,e). The spontaneous appearance of vortices following a ZFC suggests an intrinsic magnetic memory in the sample. The intrinsic magnetic field required to generate this vortex density corresponds to ~100 mG.

To find the onset temperature of this intrinsic magnetism, we repeatedly imaged the spontaneous vortex density, following thermal cycles to increasing temperature values (Figure 1c-e). The external field remained off during this process. The remnant field gradually decayed with increasing cycle temperature, until a thermal cycle reaching 3.63 K resulted in a single vortex within our field of view (Figure 1e), similar to the original state of the sample. The magnetic state therefore onsets at a temperate $T^* = 3.6$ K, with a clear separation (~1 K) between the superconducting transition temperature, and onset of magnetization.

To investigate how the remnant magnetization depends on the external field, we performed field cools, cycling the temperature to 3.5 K, slightly below $T^*$. To track the spontaneous field, we record vortex configuration formed after cooling though Tc with no external magnetic field. Therefore, we followed each FC with a ZFC (cycle temperature 2.8 K), and then imaged the vortex distribution and computed the remnant field (Figure 2a). By repeating this process with various external fields, we obtained a magnetization curve for the sample, representative of its response at 3.5 K (Figure 2b). The magnetization curve shows hysteretic behavior, reminiscent of a ferromagnet. Note that above the polarization field of the loop, the internal magnetization continues to rise linearly with external magnetic field, unlike a standard hysteresis loop. To verify that the magnetic memory is generated in the normal state, we cooled down the sample from 4.2 K to 3 K in the presence of a magnetic field, then turned the field off, and then further cooled the sample to 1.7 K (Figure 2c). In this procedure, the magnetic field was never turned on when the sample was in the superconducting state. This procedure generated the same spontaneous vortex density as the FC-ZFC procedures described above (Figure 2d,e), demonstrating that the magnetic memory is generated independently of superconductivity.

We have also verified that the magnetic memory does not decay over time in the normal state by training the sample and then keeping it at 3 K without an external field for various amounts of time (Extended Data Figure 2a). The spontaneous vortex density obtained after a subsequent ZFC to 1.7 K did not decay over time, for up to 12 minutes (Extended Data Figure 2b-d). These time scales are much larger than the electronic time scales of the system, further exclude superconducting fluctuations in the normal state as a mechanism to preserve magnetic memory.

## Absence of magnetic signals above $T_c$

The spontaneous vortex phase, magnetic memory and hysteresis above $T_c$ imply that TRS is broken at ~ 3.6 K. To determine which magnetic order exists in the system, we directly imaged the magnetic landscape of the sample both below $T_c$ (without vortices) and above $T_c$ (2.7 K < T < 3.6 K). We did not observe a signal within our magnetic noise (Figure 3). In particular, a net magnetization would generate magnetic fields at the edge of the sample. To test this, we directly imaged the magnetic landscape near the edge (Figure 3d-h). In the superconducting state, we observed a step of 1 m$\Phi_0$ across the edge (Figure 3e), due to demagnetization fields or due to interaction between the sensor and the sample[35]. In the normal state, we observed a small step, 0.1 m$\Phi_0$, across the edge (Figure 3f), corresponding to a magnetic field step of ~0.2 mG. This signal can be translated to an upper limit on the moment density. Assuming a uniform distribution of spin ½ moments, the moment density is at most ~$10^{16}$ $\mu_B$cm$^{-3}$, corresponding to an inter-moment distance of 40 nm. We also performed global magnetization measurements, which were not hysteretic in fields up to 7 T (Extended Data Figure 4) and previous studies did not detect a muon-relaxation enhancement above $T_c$ [6].

## Discussion

The magnetic field detected in the normal state is significantly (250 times) smaller than the field corresponding to the spontaneous vortex density. Our data therefore demonstrate the existence of a hidden TRSB in 4Hb-TaS2 that onsets well above the superconducting $T_c$. While the microscopic origin of this hidden order cannot be determined directly from our measurements, several mechanisms can be ruled out based on the normal state signals. It is hard to reconcile the experimental observations with conventional ferromagnetic ordering. The large inter-moment distance suggests that the interactions are dominated by Ruderman–Kittel–Kasuya–Yosida (RKKY) interactions. However, the moments are dilute compared to the electron gas, suggesting that the RKKY interactions are random in sign. This type of interactions leads to glass-like behavior, rather than ferromagnetism. In addition, for such a dilute moment concentration, Kondo screening, which is memory-free, is expected to be dominant. More importantly, in ferromagnetic superconductors[36], the spontaneous vortex density is at most the density corresponding to the magnetic field B ~ 4$\pi$M generated by the magnetization, M. In this case, the field is present in the normal state and can be detected directly using local[37–39] or global[37,40,41] probes. The lack of such signals in our samples indicates that the spontaneous vortex state cannot be caused by a conventional coupling between magnetism and superconductivity.

The proximity between the onset temperature of the hidden magnetism and the superconducting transition temperature may suggest that the TRSB state originates from the superconductivity. One trivial explanation is that the superconductor is inhomogeneous, with small areas that become superconducting at a temperature higher than bulk $T_c$. The magnetic fields generated by these areas could lead to the formation of vortices when the bulk becomes superconducting, and account for the magnetic memory. The sharp superconducting transition, characterized by multiple local and global probes (Extended data figure 1 and Refs. [6,34]) excludes this scenario.

The TRSB phase may also be a precursor to unconventional superconductivity. For example, superconducting fluctuations in a chiral superconductor could break TRS, forming a vestigial order[42,43]. Several candidate chiral superconductors were shown to produce magnetic signatures below $T_c$ [44,45], and a finite chirality could lead to a spontaneous vortex phase[46]. A chiral superconductor was previously proposed as a ground state for 4Hb-TaS$_2$, following muon spin relaxation experiments, showing an enhanced relaxation rate below $T_c$ [6]. Although our data cannot decisively confirm or exclude chiral superconductivity in 4Hb-TaS$_2$, there are several studies supporting this conclusion. Firstly, two-component superconductivity was recently reported in thin films of 2H-NbSe$_2$ [47–49], supporting a possibility for unconventional pairing channels in 4Hb-TaS$_2$. Secondly, recent scanning tunneling microscopy experiments on 4Hb-TaS$_2$ reveal edge states along layer terminations and zero-bias peaks in vortex cores, consistent with chiral superconductivity. The question is whether the proposed chiral superconducting state here generates a vestigial order. A recent study on Ba$_{1-x}$K$_x$Fe$_2$As$_2$ [50] identified two important signatures of a vestigial state: a smeared superconducting transition, observed through specific heat and resistivity, due to the order parameter fluctuations, and presence of magnetic signals in this fluctuating regime, observed through muon spin relaxation. These two signatures are absent in our case, making a vestigial order an unlikely explanation.

There are several TRSB orders not detectable by scanning SQUID microscopy. The simplest example is an antiferromagnet, which does not produce net magnetization. A more unconventional example is the TRSB state in the pseudogap phase of the cuprates[51–54]. Although the exact nature of this TRSB order parameter is under debate, it is predicted to be a loop current order where the average magnetization over a unit cell is zero[55]. Even though these orders are not detectable through SQUID measurements, they are incompatible with our results, for three reasons. Firstly, they generate a muon spin relaxation signature[56,57], which was not observed above $T_c$ in 4Hb-TaS$_2$. Secondly, they do not couple directly to an external field and therefore cannot be trained. Thirdly, since they do not generate a net magnetic flux, they cannot cause vortices in the superconductor.

One TRSB order consistent with our observations is a chiral spin liquid residing on the 1T layers. In this phase, although TRS is broken by a non-zero average spin chirality, there is little or no corresponding net magnetization. The orbital field is produced only at the third order in perturbation theory above the charge gap[58], resulting in an extremely small magnetic field. Indeed, an anomalous Hall effect was reported in the CSL candidate Pr$_2$Ir$_2$O$_7$, where no observable net magnetization was detected[59]. Thus, if the localized spins on the 1T layers do not order magnetically but form a chiral spin liquid, the spin chirality can store the magnetic memory without generating a detectable magnetic field. We note that the spin liquid phase proposed in bulk 1T-TaS$_2$ is different from the spin liquid proposed here. Namely, the bulk 1T spin liquid persists up to 200 K [23] and is likely gapless[22–24,26,28]. The CSL we propose is gapped and the chirality onsets at ~3.6 K. These differences could be driven by the interactions between the 1T and 1H layers, present in the 4Hb structure[27,28]. For example, it is possible that 4Hb hosts a gapless spin liquid at high temperatures, but it undergoes a gapping transition at a lower temperature. The resulting gapped CSL accounts for our observations in the normal state.

We now propose a mechanism through which a CSL induces spontaneous vorticity. Since the magnetic fields produced by the sample are too small to account for the spontaneous vortex density, the superconductivity and spin chirality must be coupled directly, not only through the magnetic field they produce. Such direct coupling between the spin chirality and the superconducting order parameter is possible in a chiral superconductor (CSC)[60]. The CSC generates internal magnetization which was previously shown to couple directly to magnetic dopants[61,62]. In 4Hb-TaS$_2$ the same coupling is allowed between the spin chirality of the CSL and the internal magnetization of CSC. As a result, the CSL acts as an ordering field for the CSC, aligning the CSC domains in a preferred direction and resulting in a net magnetization below $T_c$ (see Supplementary Note 1). Other chiral states residing in the 1T layer could be consistent with the data if they generate a very weak magnetization. For example, charge transfer between the 1T and 1H layers[34] could result in a chiral metallic state[33] residing on the 1T layers. We also note that our data cannot directly resolve the symmetry of the superconducting order parameter. Although a two-component order parameter provides a direct path for coupling between the chiral state and the SC, an indirect mechanism may exist for a single-component SC order parameter. For example, Kondo coupling between itinerant electrons and a chiral spin texture is expected to generate an anomalous Hall effect[63]; an analogous effect may occur in superconductors[64]. Furthermore, a magneto-elastic effect could store magnetic information in elastic deformations of the crystal, providing an alternative coupling mechanism between the hidden chiral state and the superconductor. Further work is required to provide a complete microscopic description of the effect. We finally note that the proposed chiral spin liquid and chiral superconducting natures of the 1T and 1H layers await direct confirmation.

In conclusion, we showed that the alternate stacking compound 4Hb-TaS$_2$ supports a hidden TRSB phase in its normal state. This hidden magnetic order gives rise to a spontaneous vortex phase in the superconducting state, without generating a magnetic signature in the normal state. This TRSB order is inconsistent with all classes of magnetic states previously reported to exist near superconductors – ferromagnetism, loop current orders, and vestigial chiral orders. The spontaneous vortices generated by this magnetic phase provide a new, easily accessible way to probe the coupling between unconventional magnetism and superconductivity, a puzzle shared by systems like high $T_c$ superconductors, twisted bilayer graphene, spin liquid-metal heterostructures[65,66], and recently discovered Kagome superconductors[67–69]. Future experiments on 4Hb-TaS$_2$ should determine whether the superconducting state here is indeed unconventional and study the connection between the bulk magnetic state and the topological superconductivity recently reported on the surface[34]. In other systems, our results show that Mott insulators and other frustrated magnetic states can be efficiently probed by coupling them to superconductors.

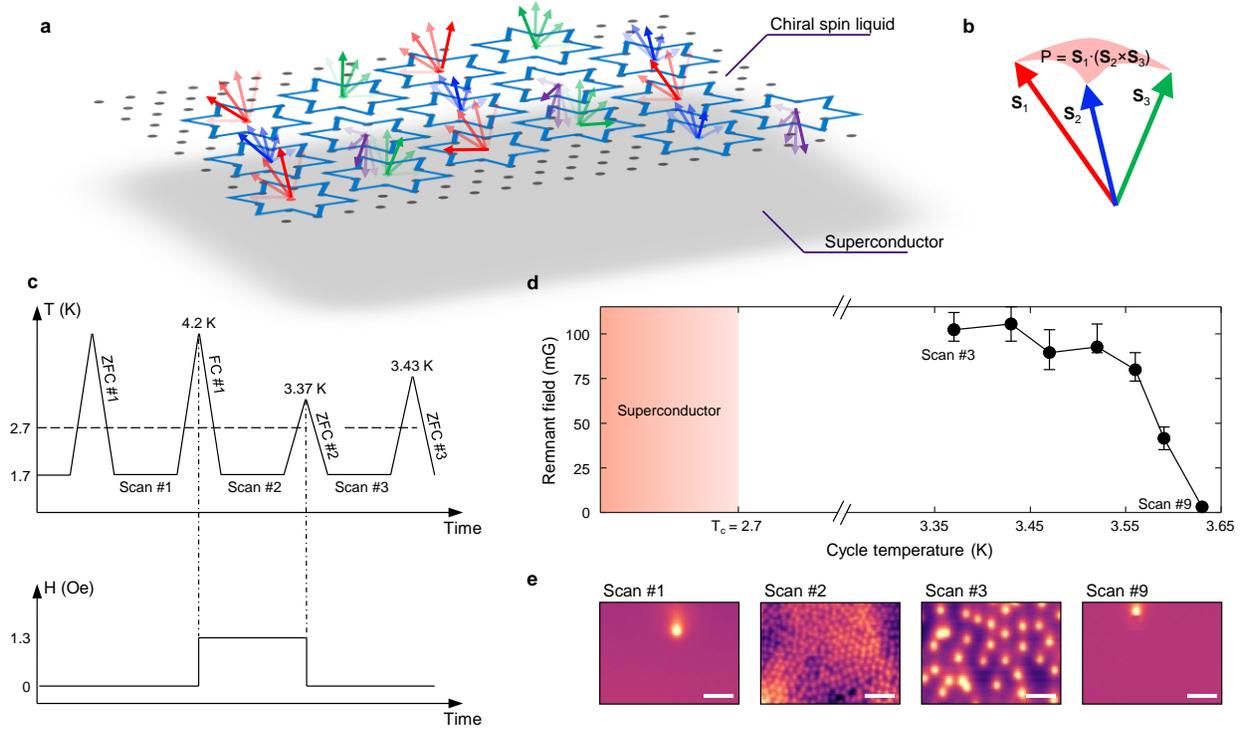

**Figure 1. Spontaneous vortex phase in 4Hb-TaS$_2$. (a)** Schematic of a 1T/1H layer composing half the 4Hb unit cell. The superconducting 1H layer interacts with the localized moments on the 1T layer through a Kondo coupling. The localized spins are arranged on a triangular lattice of stars-of-David. A chiral spin liquid residing on the 1T layers is a candidate for the TRSB state. In this phase, localized spins fluctuate but keep the chirality, shown in panel **(b)** as the solid angle between three spins, fixed. **(c)** Sample temperature (top) and external magnetic field (bottom) as a function of time, showing field cooling with varying maximum cycle temperatures. **(d)** The remnant field recorded in the sample after each zero-field-cool, as a function of the cycle temperature. The field was inferred from the spontaneous vortex density measured following each thermal cycle, at T = 1.7 K, in a 75 μm by 95 μm scan area. **(e)** SQUID images of the vortices in the superconducting state at T = 1.7 K, following each the thermal cycles. The vortices appearing in Scan#3 are spontaneous. Scale bars, 20 μm.

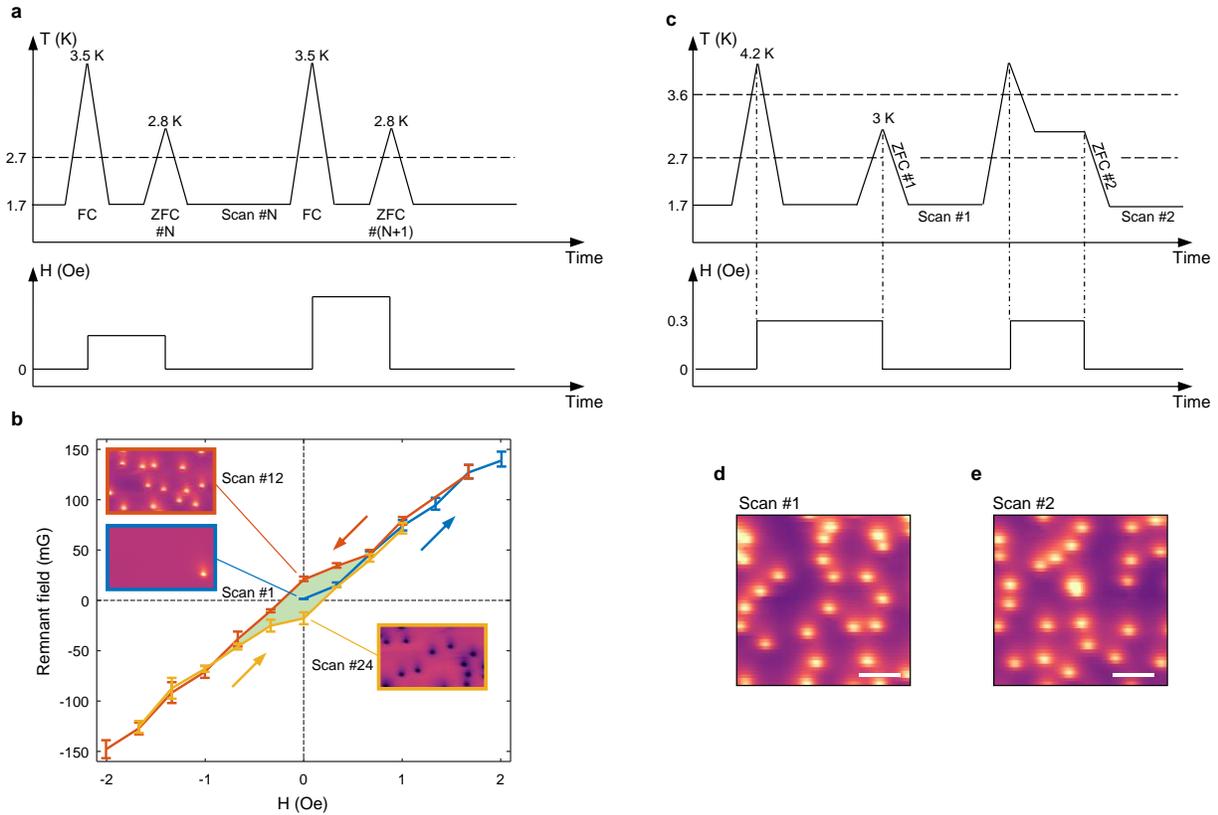

**Figure 2. Magnetic hysteresis. (a)** We repeatedly cycled the temperature to 3.5K and cooled down in the presence of various fields, H. To separate the spontaneous vortices, we followed with ZFC (cycle temperature 2.8 K) and mapped the vortex distributions. **(b)** Each point represents the remnant field determined from the vortex distribution, following the thermal cycling procedure described in a. The x axis is the external field applied during the FC stage, starting with H = 0, to 2 G (blue line) to –2 G (orange) to 1 G (yellow). Insets: magnetic flux images of vortex distribution at H = 0, during the various stages of the magnetization curve. **(c)** Two field-training protocols which generated the same spontaneous vortex density. In the first protocol the magnetic field was turned on upon the first cooldown through $T_c$. In the second protocol, the sample was trained by field-cooling it through the TRSB transition, but the field was turned off before entering the superconducting state. **(d, e)** SQUID images of the spontaneous vortices in the superconducting phase following the **(d)** first protocol and the **(e)** second protocol. Scale bars, 30 μm.

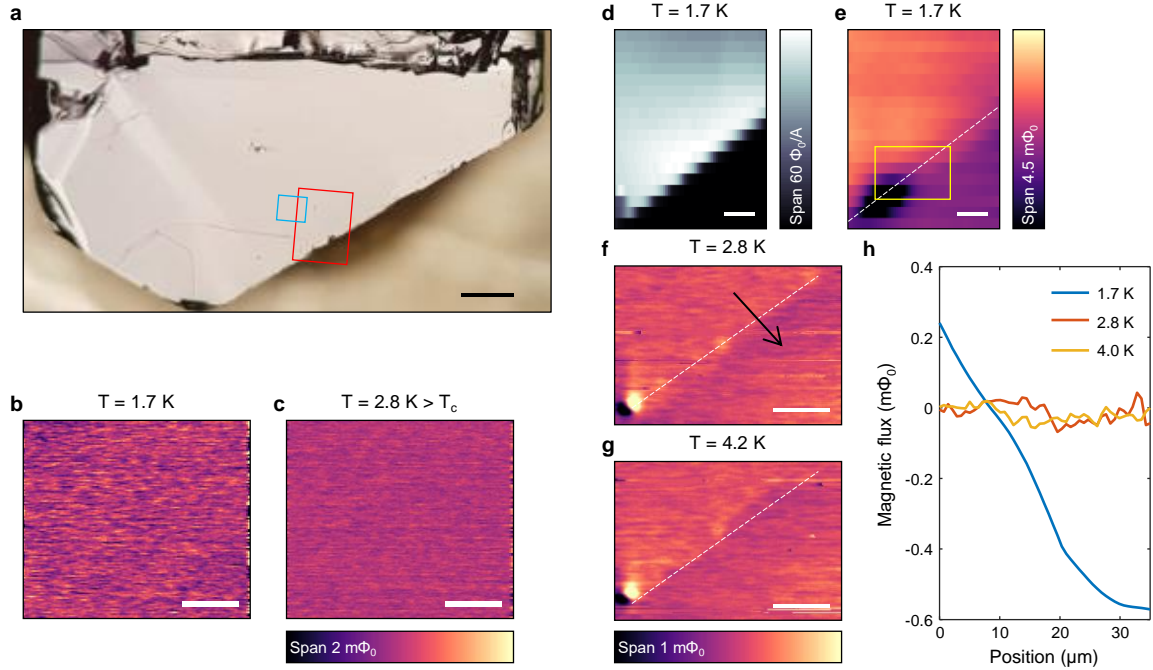

**Figure 3. Absence of magnetic signals in the normal state. (a)** Optical microscope image of the sample. Scale bar, 150 µm. **(b,c)** SQUID magnetometry images of the blue rectangle in a, taken at 1.7 K **(b)** and at 2.8 K **(c)**. The data were taken after cooling down the sample from 4 K, in the absence of an external magnetic field (< 5 mG). Within our noise level, there are no spatial variations neither below nor above Tc, within this field of view. Scale bars, 20 µm. **(d,e)** Susceptibility **(d)** and magnetometry **(e)** maps of the region marked by a red rectangle in a, taken at 1.7 K. The susceptibility map shows a sharp change in the diamagnetic response, clearly identifying the edge of the sample. The small magnetic signal **(e)** below $T_c$ is due to demagnetization effects or interactions between the sample and the sensor. Scale bars, 30 µm. **(f,g)** Magnetometry images of the yellow rectangle in c, taken at 2.8 K **(f)** and 4 K **(g)**. The images show a faint magnetic signal (0.1 m$\Phi_0$) at the edge of the sample. This signal cannot reconcile a homogeneous remnant field of more than 1 mG, and does not vanish above 3.6 K. The dashed lines in panels e-g mark the edge of the sample. Scale bars, 20 µm. **(h)** Line cuts from panels e-g, taken along the black arrow in panel f, perpendicular to the sample's edge.

# Methods

## Sample preparation

High-quality single-crystals of 4Hb-TaS$_2$ were grown using the chemical vapor transport method. A stoichiometric mix of Ta and S was sealed in a quartz ampule under vacuum. Addition of 1% Se significantly improves sample quality. The mixture underwent a sintering process, forming a boule of 4Hb-TaS$_{1.99}$Se$_{0.01}$. The boule was crashed and placed in a 200-220 mm long quartz ampule with a 16 mm diameter. Iodine was added as a transport agent, and the ampule was sealed under vacuum. The ampoule was then placed in a 3-zone furnace, where the hot ends were heated to 800 °C and the middle part was kept at 750 °C. After about 30 days the ampoule was quenched in cold water.

The crystal structure and chemical composition were verified[6] using X-ray diffraction (XRD) and electron energy dispersive spectroscopy (EDS) in a scanning electron microscope (SEM).

## Scanning SQUID microscopy

Simultaneous local magnetometry and susceptometry measurements were performed[70,71] by scanning a planar SQUID with a 0.75 µm pickup loop over a cleaved surface of the sample. Susceptometry measurements were performed by a mutual inductance measurement using a second coil (diameter 3 µm) concentric to the SQUID's pickup loop. An AC current (1 mA RMS, frequency ~ 1 kHz) applied to this coil generates a local magnetic field of ~1.5 G RMS. The superconductor screens this field, generating a change in the flux penetrating the SQUID's pickup loop. This change was recorded simultaneously to the DC measurement using a lock-in amplifier. Scans were performed at a constant sensor-sample distance.

External magnetic field was applied using a copper coil wrapped around the microscope setup. Earth's magnetic field was reduced by including two low-temperature µ-metal (Amumetal 4K, Amuneal) cylindrical shields with diameters of 96 mm and 100 mm around the microscope and coil, achieving fields as small as 5-10 mG near the sample. Further cancelation was achieved by applying a small current through the copper coil, reaching fields ~ 3 mG. "Zero field" measurements were performed under these conditions.

Temperature was monitored with a Si diode and thermal cycles were performed at a rate of ~0.5 K/min with an open loop control. The peak temperature at each cycle was held for 3 minutes before the sample was cooled down. Before thermal cycles, the SQUID was retracted 100 µm from the sample to avoid field distortions due to the superconducting lines on the sensor.

## NbSe$_2$ flake as a probe for stray magnetic fields

To eliminate magnetic field sources in our system as the source of the magnetic memory and thus verify that the magnetic memory originates from the 4Hb-TaS$_2$ itself, we performed the following control experiment. We used the Meissner response of a NbSe$_2$ flake, recorded by the SQUID, as a probe for ambient magnetic fields (Extended Data Figure 3a,b). We replicated the field-training process of Figure 1 by cooling down the NbSe$_2$ flake from 4.2 K to 1.7 K with an external field of 0.25 Oe. We then imaged the Meissner response at 1.7 K before and after removing the external field (Extended Data Figure 3c,d). The zero field images at both 1.7 K and 4.2 K reveal no Meissner response, suggesting that no additional magnetic fields were created in the vicinity of the flake following the "training" process. The 4Hb-TaS$_2$ data (Figure 2) shows that the remnant fields following field cooling through 3.6 K are ~10% of the external field applied during the cooling. The data show that the Meissner response after the field is turned off

(Extended Data Figure 3e) is much smaller than the expected training effect. The absence of memory in the NbSe$_2$ sample suggests that the effect observed in 4Hb-TaS$_2$ is intrinsic, and not related to elements of the scanning SQUID setup.

### Global magnetometry and resistivity measurements

Global magnetization measurements were performed using a SQUID magnetometer (MPMS, Quantum Design). A ~10 mg sample was glued to a quartz sample holder using GE varnish. For determining T$_c$, the sample was cooled at zero field to 1.8 K. The magnetization was then measured as a function of temperature between 1.8 K and 5 K, with an external field of 100 mG. To measure the magnetization at the normal state, the sample was cooled at zero field to 2.8K. The magnetization was measured while the field is ramped to 7 T, then swept to -7 T and back to zero field.

Resistivity measurements as a function of the temperature were performed in a commercial system (Quantum Design DynaCool PPMS). Electrical contacts in a Van der Pauw configuration were made using silver paste. The IV curves were verified to be linear in the current range used for the measurements.

## Method references

## Data availability
The data that support the findings of this study are available from the corresponding author upon reasonable request.

## Acknowledgments
We thank H. Beidenkopf, Y. Dagan, A. Aurbach, E. Shimshoni, R. Ilan and F. de Juan for fruitful discussions. We thank T.R. Devidas for assistance with the $NbSe_2$ measurements. E.P., A.V.B. and B.K. were supported by the European Research Council Grant No. ERC-2019-COG-866236, the Israeli Science Foundation grant no. ISF-1281/17, COST Action CA16218, and the QuantERAERA-NET Cofund in Quantum Technologies, Project No. 731473. E.B. was supported by the European Research Council Grant No. ERC-2019-COG-817799. J.R. was supported by the Israeli Science Foundation grant no. ISF-994/19. A.K. was supported by Israeli Science Foundation grant no. ISF-1263/21.

## Author contributions
E.P and B.K designed the experiments. E.P, A.V.B and B.K performed the scanning SQUID measurements. I.F and A.A prepared the samples. A.A performed the global characterization measurements. E.P, B.K, E.A, E.B, I.K, J.R and A.K discussed the data and interpreted the results. E.P, J.R and B.K wrote the manuscript with contributions from all coauthors.

## Competing interests
The authors declare no competing interests.

## Additional information
**Supplemental information** contains supplemental discussion.

**Correspondence and requests for materials** should be addressed to eylon.persky@biu.ac.il, beena@biu.ac.il

# Extended data

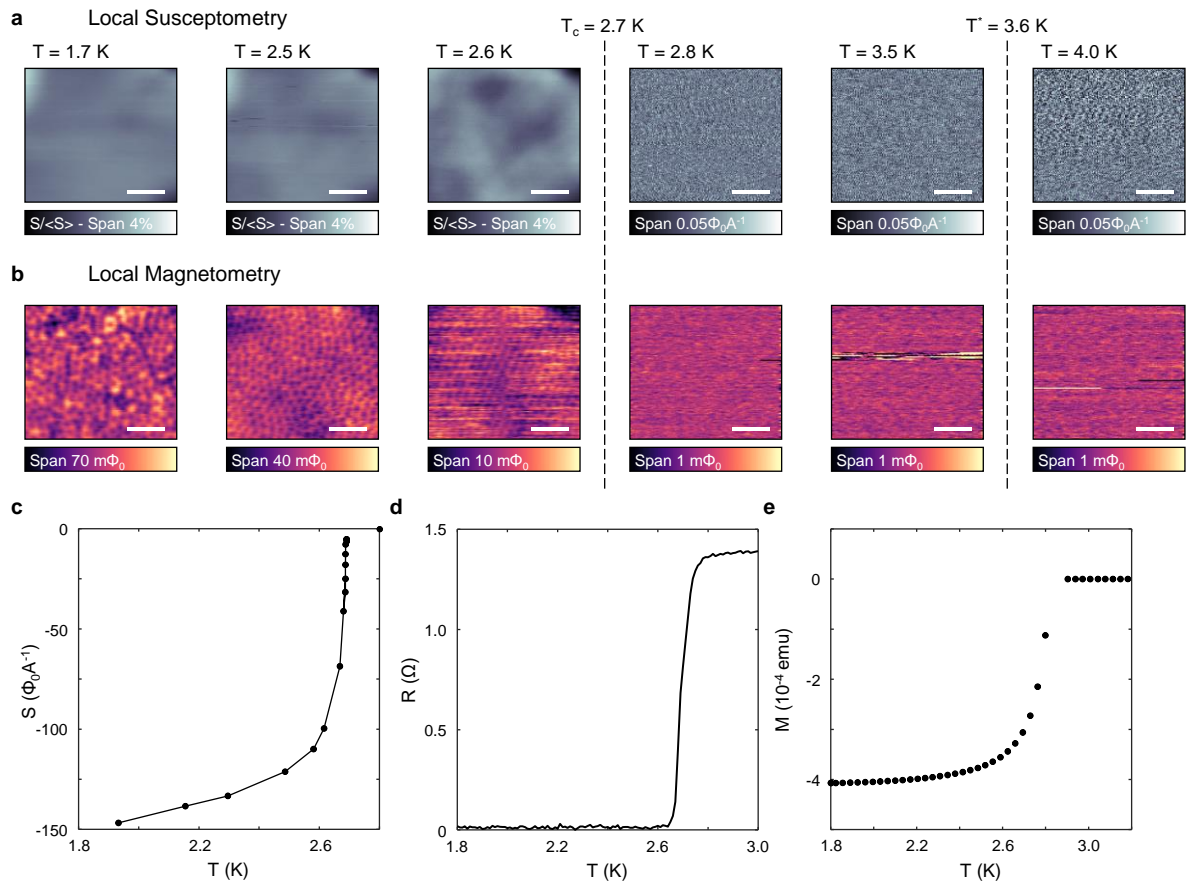

**Extended Data Figure 1. Determination of $T_c$.** **(a)** Local susceptibility maps taken at different temperatures below and above $T_c$. Below $T_c$, the diamagnetic response is homogeneous to within ±2% of the space-averaged signal, even at T = 2.6 K, close to $T_c$. At T = 2.8 K (>$T_c$) we detect no signal within our noise level, demonstrating that the system is completely normal within our field of view. **(b)** Local magnetometry images of the sample following a field cool at various temperatures. The images clearly show vortices below $T_c$, which are completely absent above $T_c$ (note the change to the color span at T = 2.8 K). Scale bars, 20 μm. **(c)** Local temperature dependence of the susceptibility taken at a representative single point on the sample. **(d)** Resistance measurements as a function of temperature. **(e)** Global magnetization measured after field cooling the sample with an external field of 100 Oe. All measurements show a sharp superconducting transition at ~ 2.7 K.

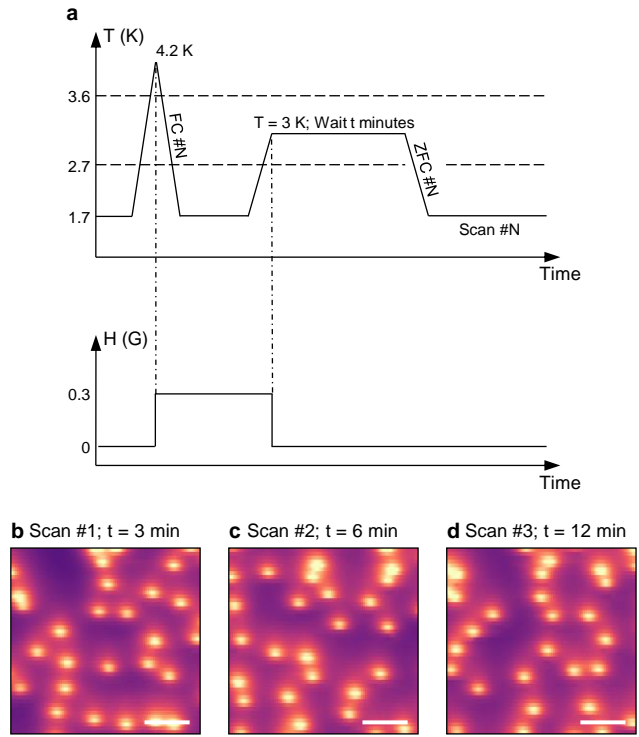

**Extended Data Figure 2. Absence of magnetic memory decay in the normal state. (a)** The sample was first trained by field-cooling it through 3.6 K. It was then kept at 3 K for various amounts of time, before ZFC to 1.7 K and measuring the resulting vortex density. **(b-d)** SQUID images of the spontaneous vortices in the superconducting phase after waiting for **(b)** 3 minutes, **(c)** 6 minutes, and **(d)** 12 minutes at 3 K. The vortex density did not change as a function of time waited at the normal state. Scale bars, 30 μm.

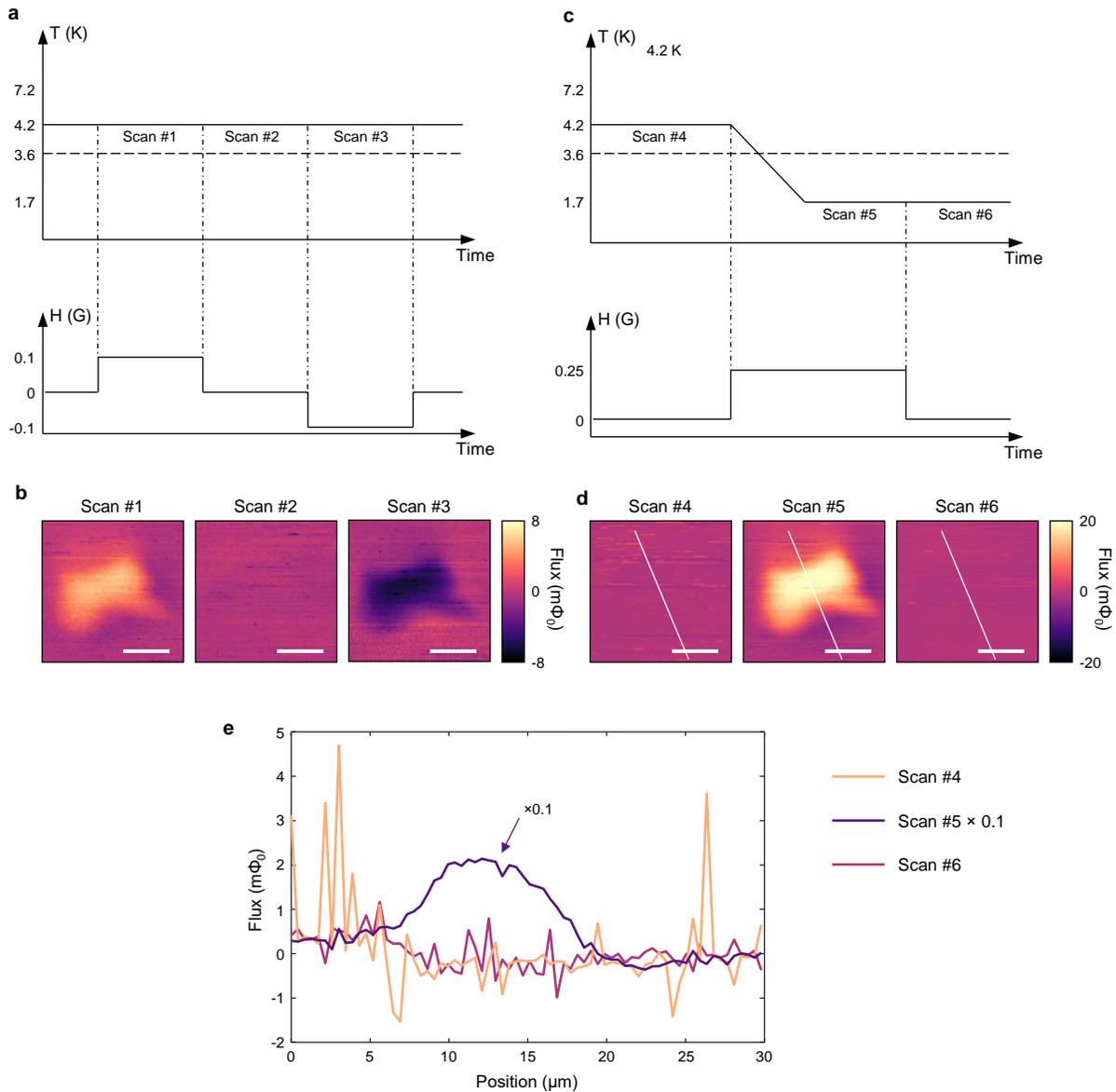

**Extended Data Figure 3. Absence of magnetic memory in a NbSe₂ flake. (a)** To demonstrate how the Meissner response of a NbSe2 flake can be used to probe external magnetic fields, we measured its magnetic signal at 4.2 K, at various magnetic fields. **(b)** The corresponding magnetic flux images at 4.2 K, showing signals due to the Meissner response. Note that both the presence of an external field and its polarity can be detected through the Meissner effect. When the field is turned off (scan #2) the signal disappears. **(c)** A "field cooling" protocol from 4.2 K to 1.7 K, like that used in Figure 1. **(d)** The corresponding magnetic flux images. After the field is turned off at 1.7 K (scan #6), the Meissner response disappears, demonstrating that there is no remnant field in the system. **(e)** Line cuts showing the Meissner response from scans #4-6. The data from scan #5 (field on) is multiplied by 0.1.

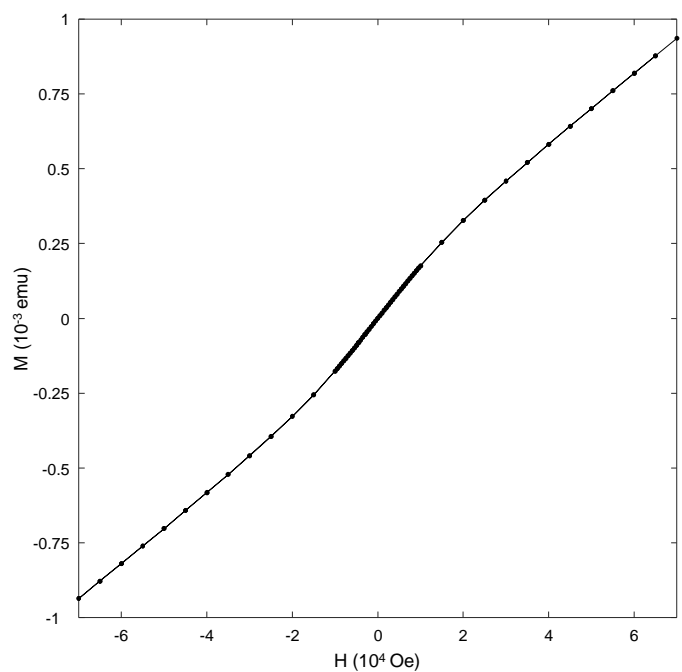

**Extended Data Figure 4. Absence of global magnetic signal above T$_c$.** The global magnetization of the sample as a function of the external field, taken at 3.2 K. The magnetic response was not hysteretic for fields up to 7 T.

# Supplementary Information for
# Magnetic memory and spontaneous vortices in a van der Waals superconductor

Eylon Persky[1,2], Anders V. Bjørlig[1,2], Irena Feldman[3], Avior Almoalem[3], Ehud Altman[4,5], Erez Berg[6], Itamar Kimchi[7], Jonathan Ruhman[1], Amit Kanigel[3] and Beena Kalisky[1,2]

1. Department of Physics, Bar Ilan University, Ramat Gan 5920002, Israel
2. Bar Ilan Institute of Nanotechnology and Advanced Materials, Bar Ilan University, Ramat Gan 5920002, Israel
3. Department of Physics, Technion–Israel Institute of Technology, Haifa 32000, Israel
4. Department of Physics, University of California, Berkeley, California 94720, USA
5. Materials Sciences Division, Lawrence Berkeley National Laboratory, Berkeley, California 94720, USA
6. Department of Condensed Matter Physics, Weizmann Institute of Science, Rehovot 76100, Israel
7. School of Physics, Georgia Institute of Technology, Atlanta, Georgia 30332, USA

Supplementary Note 1. Ginzburg Landau theory for the hidden magnetization

Here we provide a phenomenological Ginzburg-Landau theory in two spatial dimensions to describe the coupling between the chiral spin liquid assumed to reside on the 1T layers and the superconducting state on the H layers.

The first element in the theory is the chiral state residing on the T layers, which is captured by the free energy density

$$f_{SL} = a_P(T - T_*)|\mathbf{P}|^2 + b_P|\mathbf{P}|^4 - \mu\, \mathbf{P} \cdot \mathbf{B}, \tag{1}$$

where $\mathbf{P} = \mathbf{S}_1 \cdot (\mathbf{S}_2 \times \mathbf{S}_3)\hat{\mathbf{z}}$ is the chiral order parameter, $\mathbf{S}_1, \mathbf{S}_2$ and $\mathbf{S}_3$ are the three spin operators on the plaquette of the triangular lattice of stars-of-David and $T_*$ is the ordering temperature. The last term, which is proportional to µ, is the coupling of the chiral order parameter to the magnetic field $\mathbf{B}$. Following Ref. [1] we have

$$\mu = \frac{\pi A_\Delta}{18\Phi_0} \frac{W^3}{U^2},$$

where $A_\Delta = 13a^2 \sin \pi/3$ is the area of the super unit-cell, a = 3.38 Å, $\Phi_0$ is the flux quantum, W ≈ 50-100 meV is the width of the flat miniband[2], and U ≈ 200 meV is the on-site interaction strength. The magnetic field generated by a finite chirality, $\langle P \rangle \neq 0$ (which we assume to be of order one), is then given by $|\mathbf{B}_P| = 4\pi\mu|\mathbf{P}| \approx 13$ pT. Thus, the generated field is expected to be much below the sensitivity of our probe.

For the superconducting state, we consider two order parameters. As explained below, the enhanced magnetic field corresponding to the spontaneous vortex density can be explained with the internal magnetization of a chiral superconductor. However, since bulk 2H-TaS$_2$ is an s-wave superconductor[3], it is necessary to include a single-component order parameter in the theory[4]. This s-wave order parameter, Δ, is described by the free-energy density

$$f_\Delta = a_\Delta \left(T - T_c^{(1)}\right) |\Delta|^2 + b_\Delta |\Delta|^4 + \frac{1}{2m^*} |\mathbf{D}\Delta|^2, \tag{2}$$

where $T_c^{(1)}$ is the s-wave transition temperature, $\mathbf{D} = -i\nabla - 2e\mathbf{A}$ is the gauge invariant derivative and $\mathbf{A}$ is the vector potential related to the magnetic field through $\mathbf{B} = \nabla \times \mathbf{A}$. The second order parameter belongs to one of the two-component representations of the trigonal point group symmetry D3, which can therefore support a chiral superconducting state and is described by the free-energy density[4–7]

$$f_\eta = a_\eta \left(T - T_c^{(2)}\right) |\boldsymbol{\eta}|^2 + b_\eta |\boldsymbol{\eta}|^4 + c_\eta |\boldsymbol{\eta}^* \times \boldsymbol{\eta}|^2 + \kappa_1 \left[|\mathbf{D}\eta_x|^2 + |\mathbf{D}\eta_y|^2\right] + \kappa_2 [|\mathbf{D} \cdot \boldsymbol{\eta}|^2 - |\mathbf{D} \times \boldsymbol{\eta}|^2], \tag{3}$$

where $\boldsymbol{\eta} = (\eta_x, \eta_y)$ is a vector of the two components of the order parameter. For $c_\eta < 0$ the system prefers a chiral ground state with $i\langle \boldsymbol{\eta}^* \times \boldsymbol{\eta} \rangle \neq 0$.

Integrating the last term by parts leads to the following contribution to the free-energy

$$i \kappa_2 (\boldsymbol{\eta}^* \times \boldsymbol{\eta}) \cdot \mathbf{B},$$

which remains finite in the uniform chiral state. Therefore, a uniform chiral state generates a magnetic field in the z direction

$$\mathbf{B}_\eta = 4\pi \kappa_2 \langle i\boldsymbol{\eta}^* \times \boldsymbol{\eta} \rangle. \tag{4}$$

It is important to note that without an external ordering field the SC chiral order parameter $i\boldsymbol{\eta}^* \times \boldsymbol{\eta}$ is expected to be break into domains of opposite chirality[8], $\boldsymbol{\eta}_\pm = \eta_0(1, \pm i)$. The net magnetic field is then an average over these domains.

The superconductor does not sustain the field in Eq. (4) when it is non-zero. If $B_\eta < H_{c1}$ it will screen the magnetic field by means of Meissner currents on domain or sample boundaries. On the other hand, for $B_\eta > H_{c1}$ one expects the magnetic field to penetrate in the form of vortices. Such vortices are not generated by an external field and may appear even when an external field is absent, if the domain structure is imbalanced and has a net magnetization. Ng and Varma[9] have discussed this scenario in the context of ferromagnetic materials that become superconducting. The situation here is similar[10], where the SC chiral order parameter takes the role of the magnetization of the ferromagnet.

The final element in the model is the coupling between the three fields $\mathbf{P}, \Delta$ and $\boldsymbol{\eta}$. By symmetry, the allowed couplings are

$$f_{coupling} = \lambda_{\Delta\eta}|\Delta|^2|\boldsymbol{\eta}^* \times \boldsymbol{\eta}|^2 + \lambda_{\Delta P}|\Delta|^2|\mathbf{P}|^2 + \lambda_{P\eta}(i\boldsymbol{\eta}^* \times \boldsymbol{\eta}) \cdot \mathbf{P} \qquad (5)$$

Note that the third term in Eq. (5) is linear in $\mathbf{P}$ and is thus the only term that allows the domain structure embedded in the chiral spin-liquid to be transmitted to the superconducting state as described below. We may consider two scenarios to explain the spontaneous vortex state:

- *The superconducting order parameter is chiral* – In this case the coupling to the chiral order parameter $\langle \mathbf{P} \rangle$ acts as an effective magnetic field for the SC chiral order, $i\langle \boldsymbol{\eta}^* \times \boldsymbol{\eta} \rangle = \frac{\lambda_{P\eta}}{2c_\eta}\langle \mathbf{P} \rangle$. If $c_\eta \propto T - T_c^{(2)}$ is divergent at the transition point (equivalent to assuming that there a single transition into the chiral superconducting state without a vestigial order), the chiral spin-liquid acts as an ordering field for the SC chiral order parameter. In turn, the CSC generates a spontaneous internal magnetization proportional to $\langle \mathbf{P} \rangle$, $\mathbf{B}_\eta = 4\pi\kappa_2\lambda_{P\eta}/c_\eta\langle \mathbf{P} \rangle$. If $B_\eta > H_{c1}$ this leads to a spontaneous vortex state.
- *The superconducting state is s-wave* – In this case there is no direct coupling of the chiral order parameter $\mathbf{P}$ to the superconducting one $\Delta$. However, we may still assume that $c_\eta$ is small and that $\lambda_{\Delta\eta} < 0$, which implies that the chiral order is soft, and that the chiral and s-wave orders are cooperative and not competing. In this case, when the superconducting order parameter condenses ($\langle\Delta\rangle \neq 0$), the susceptibility of $i\langle\boldsymbol{\eta}^* \times \boldsymbol{\eta}\rangle$ is shifted, $2/c_\eta \to 2/(c_\eta - \lambda_{\Delta\eta}\langle|\Delta|^2\rangle)$. In this way the superconducting transition into the s-wave state can also push the TRSB order parameter to develop a finite expectation value $i\langle\boldsymbol{\eta}^* \times \boldsymbol{\eta}\rangle$, or enhance it if it existed prior to the transition. This resembles the vestigial phase predicted to exist above the transition temperature of nematic and chiral superconductors[11], where the order parameter itself $\boldsymbol{\eta}$ does not condense, while its composite does.

Note that the order parameter P can represent TRSB states other than a chiral spin liquid, provided that they produce a weak magnetic field (support small µ). For example, a chiral metal is expected to generate a magnetization significantly smaller than 1 $\mu_B$ per site. Such state could possibly exist in the 1T layers as a result of charge-doping the spin liquid state[12]. In both cases, coupling to a chiral superconductor provides a direct mechanism for generating a spontaneous vortex phase.